\documentclass[a4paper]{jpconf}

\usepackage{graphicx}
\usepackage{subfigure}
\usepackage{amsmath}
\usepackage{amssymb}
\usepackage{enumerate}
\usepackage{psfrag}
\usepackage{bm}
\usepackage{color}
\usepackage[utf8]{inputenc}
\usepackage[square,sort&compress]{natbib}

\newcommand{\overlap}[2]{{\ensuremath \bigl \langle \,  #1 \, \bigr | \,  #2  \, \bigr \rangle }}
\newcommand{\braket}[3]{{\ensuremath \bigl \langle \,  #1 \, \bigr | \, #2 \, \bigl | \,  #3  \, \bigr \rangle }}
\newcommand{\bigbraket}[3]{{\ensuremath \Bigl \langle \,  #1 \, \Bigr | \, #2 \, \Bigl | \,  #3  \, \Bigr \rangle }}
\newcommand{\ket}[1]{{\ensuremath  \bigl | \,  #1  \, \bigr \rangle }}

\newcommand{\smallket}[1]{{\ensuremath  | \,  #1  \,  \rangle }}
\newcommand{\bra}[1]{{\ensuremath \bigl \langle \,  #1 \, \bigr | }}
\newcommand{\smallbra}[1]{{\ensuremath \langle \,  #1 \,  | }}

\newcommand{\tderiv}{\ensuremath \frac \partial{\partial t}}

\begin{document}

\title{Multiconfigurational time-dependent Hartree-Fock calculations for photoionization of one-dimensional Helium}

\author{David Hochstuhl, Sebastian Bauch, Michael Bonitz}
\address{Institut f\"ur Theoretische Physik und Astrophysik, D-24098 Kiel, Germany}
\date{\today}

\begin{abstract}
The multiconfigurational time-dependent Hartree-Fock equations are discussed and solved for a one-dimensional model of the Helium atom. Results for the ground state energy and two-particle density as well as the absorption spectrum are presented and compared to direct solutions of the time-dependent Schrödinger equation.
\end{abstract}



\section{Introduction}

In recent years new powerful radiation sources became available for the precise investigation of photoionization processes of matter. New methods have made it possible to observe the electronic motion in a time-resolved fashion on a scale of attoseconds \cite{Krausz_2009,Brabec_2000}. With this, several effects, such as strong-field tunneling \cite{Uiberacker_2007} or time-resolved Auger decay \cite{Drescher_2002} could be studied in detail. The explanation of the arising effects is a challenge for theoretical physicists, which need to face an old problem, namely the solution of the electronic Schrödinger equation with its exponentially growing effort with increasing number of the degrees of freedom.\\
Several methods have been developed to circumvent this fundamental limitation. Among them are time-dependent density functional theory (TDDFT), see e.g.~\cite{Runge_1984}, the method of nonequilibrium Green functions (NEGF), e.g.~\cite{Kadanoff_Baym, Keldysh_1964, Dahlen_2007}, or time-dependent reduced density-matrix theory (TDRDM), e.g.~\cite{Bonitz_QKT,Kremp_1997,Viktor_1995}, which all aim at projecting the Schrödinger equation on a more convenient set of equations requiring only a polynomially growing effort in solution. Despite the indisputable successes of these methods, they lack a systematical and practically feasible way to achieve convergence to the exact result.\footnote{In TDDFT, this is actually the main problem, since the result crucially depends on the choice of the exchange-correlation functional. NEGF (and TDRDM) are principally exact, if all equations in a hierarchy of equations were taken into account, resp. if all self-energy diagrams were summed up. In practice, however, the hierarchy is decoupled already on a low level.}

In this paper, we apply a method which provides this mentioned feature, namely time-dependent Multiconfigurational Hartree-Fock (MCTDHF). It can either be seen as an extension of Hartree-Fock, which includes several Slater determinants (or permanents in the case of Bosons) instead of a single one, or as an extension of Configuration Interaction, that employs time-dependent single-particle orbitals instead of a fixed basis. In the case of infinitely many determinants, the results essentially become exact. In MCTDHF, the exponential problem is not really avoided, but it is postponed to much larger systems than in direct solutions of the Schrödinger equation. Thus, MCTDHF is applicable to few particle systems of roughly ten particles. To become familiar with the method, in this work we consider a directly solvable one-dimensional model of the Helium atom, and compare the MCTDHF results with those from the time-dependent Schrödinger equation (TDSE).

The outline of the paper is as follows: After this introduction, we give an overview on the MCTDHF formalism, recapitulate the working equations and provide the main ideas of our implementation. Subsequently, groundstate as well as time-dependent results for one-dimensional Helium are presented. In this paper, we apply the notation of Ref.~\cite{Helgaker}.

\section{The MCTDHF method}
The system of our interest are few electron atoms in an external electromagnetic field described by the Hamiltonian (in atomic units)
\begin{align}
\label{Hamiltonian} \hat H \ = \ \sum_{k=1}^N \, \Biggl\{ \frac{{\mathbf p_k}^2}2 \, - \, \frac{Z}{|\mathbf r_i|} \, - \, \boldsymbol {\mathcal E}(t)\cdot \mathbf r_k \Biggr\} \ + \ \frac12 \, \sum_{k\neq l} \, \frac{1}{|\mathbf r_k-\mathbf r_l|}\,,
\end{align}
where the motion of the nucleus is neglected. In this paper, we focus on the one-dimensional Helium atom ($Z=2$), for which the singularities in the Coulomb potential are softened by a cutoff, see Sec.~\ref{sec:helium_model}. However, since the following theoretical considerations are completely general, we use the three-dimensional formulation. The studied systems are assumed to be initially in the groundstate, so a wavefunction treatment is appropriate.

\subsection{Overview}
Standard wavepacket propagation methods, e.g. time-dependent Configuration Interaction, typically approximate the many-body wavefunction as a linear expansion in a set of basis vectors of the subspace  $\mathcal H(2M,N)$ of the $N$-particle Hilbert space, that is, the subspace spanned by all $N$-fold anti-symmetrized products of $2M$ single-particle orbitals (the factor $2$ is due to the two possible spin-projections). Commonly, the many-body basis states are taken to be either Slater determinants (SD) or configuration state functions (CSF). The latter are special linear combinations of Slater determinants, which are not only eigenfunctions of the projected-spin operator $\hat S_z$ -- as Slater determinants are -- but also eigenfunctions of the total-spin operator $\hat S^2$ \cite{Helgaker}. This ansatz for the wavefunction is inserted into the Schrödinger equation to obtain a linear equation of motion for the time-dependent expansion coefficients, which may be solved by a couple of methods (e.g. short iterative Lanczos- \cite{Park_1986, Jie_2006}, Chebycheff- \cite{Tal-Ezer_1984}, split operator- \cite{Chin_2001} methods, etc. \cite{Leforestier_1991}). However, this simple approach suffers from a main drawback, namely that the size of the N-particle Hilbert space basis grows exponentially with the number of particles $N$ and orbitals $2M$, what restricts the applicability to rather small systems. A common way around this problem is to drop several Slater-determinants, which are believed to be physically less important, as it is done for instance in the time-dependent Configuration Interaction singles (TDCIS) method \cite{Rohringer_2006}. This enables the treatment of higher particle numbers resp. the inclusion of a larger single-particle basis at the cost of a reduced description of correlation effects.

The multiconfigurational time-dependent Hartree-Fock method uses an alternative strategy. It also approximates the wavefunction by a linear expansion of basis states of $\mathcal H(2M,N)$, which are, however, allowed to vary in time. This is achieved by assuming the $2M$ orbitals $\{ \smallket{\phi_k} \}$ to be explicitly time-dependent, and being expressed by an expansion in a set of $N_b$ time-independent orbitals $\{ \smallket{\chi_l} \}$:
\begin{align}
\label{basis_ex} \ket{ \phi_k } (t) \ = \ \sum_{l=1}^{N_b} b_{kl} \,(t) \; \ket{\chi_l} \, , \quad k=1,\cdots,2M \,.
\end{align}
By using Slater determinants built with such time-dependent orbitals, it is likewise only possible to represent states in a subspace $\mathcal H(2M,N)$. However, now this space is allowed to vary freely in the much larger subspace $\mathcal H (2 N_b,N)$. In this way it is possible to arrive at an accurate description of the wavefunction and to defer the problem of the exponential growth of the Hilbert basis size to larger systems.\\
In the following, we give the MCTDHF equations for systems described by the standard electronic Hamiltonian (\ref{Hamiltonian}), which in second quantization reads
\begin{align}
\hat H (t) \ = \ \sum_{pq} \, h_{pq}(t) \, \hat  E_{pq} + \frac 12 \sum_{pqrs} \, g_{pqrs} \, \hat  e_{pqrs}\,,
\end{align}
with the one- and two-particle excitation operators \cite{Helgaker}
\begin{align}
\hat E_{pq} \ &= \ \sum_\sigma \, \hat  a^\dagger_{p\sigma} \hat  a_{q\sigma}\,,\\
\hat e_{pqrs} \ &= \ \sum_{\sigma\tau} \, \hat a^\dagger_{p\sigma} \hat a^\dagger_{r\tau} \hat a_{s\tau} \hat a_{q\sigma}\,,
\end{align}
acting in the time-dependent basis $\{ \smallket{\phi_l} \}$. The electron integrals are given by
\begin{align}
h_{pq}(t)  &=  \int d \mathbf r \, \phi^\ast_{p}(\mathbf r) \left\{-\frac 12 \Delta + V(\mathbf r) - \boldsymbol {\mathcal E}(t) \cdot \mathbf r \right\} \phi_{q}(\mathbf r)\,,\\
g_{pqrs}  &=  \iint d\mathbf r \, d \mathbf {\bar r} \; \phi^\ast_{p}(\mathbf r) \phi_{q}(\mathbf r) \, \frac1{|\mathbf r -\mathbf {\bar r}|} \, \phi^\ast_{r}(\mathbf {\bar r}) \phi_{s}(\bar {\mathbf r})\,.
\end{align}
The single particle Hamiltonian $\hat h(t)$ includes the action of an external electromagnetic field $\boldsymbol {\mathcal E}(t)$ in dipole approximation (length gauge) and thus carries the only explicit time-dependence.

\subsection{The MCTDHF equations}
As mentioned above, the multiconfigurational time-dependent Hartree-Fock ansatz approximates the wavefunction by a linear combination of time-dependent basis states of the $N$-particle Hilbert space, which are in the following assumed to be Slater determinants:
\begin{align}
\ket{\Psi} \ = \ \sum_{\mathbf n} \; C_{\mathbf n}(t) \; \ket{n_{1\alpha} , n_{1\beta} , n_{2\alpha} \cdots , n_{M \beta} ;t}\,.
\end{align}
The Slater determinants are written in occupation number representation specifying the occupation of the $M$ time-dependent spatial orbitals $\{ \smallket{\phi_k} \}$ with an electron with spin-projection $\alpha$ (spin-up) or $\beta$ (spin-down), and $\sum_k n_{k\alpha} +n_{k\beta} \equiv N$. This corresponds to a \emph{spin-restricted} treatment, i.e. we assume that $\alpha$- and $\beta$-electrons share a common spatial orbital \cite{Hochstuhl_2010}.\\
For the derivation of the equations of motion we follow Refs.~\cite{Cederbaum_boson,Cederbaum_unified} and employ the Lagrange formulation of the time-dependent variational principle, in which the action functional
\begin{align}
& S\Bigl[\bigl \{C_{\mathbf n}(t) \bigr \}, \bigl \{\smallket{\phi_k}(t)\bigr\} \Bigr] \ = \ \int dt \ \Bigg \{ \bigbraket{\Psi}{\hat H - i\frac{\partial}{\partial t}}{\Psi}
 -  \, \sum_{kl} \mu_{kl}(t) \, \Bigl( \overlap{\phi_k}{\phi_l} -\delta_{kl} \Bigr) \Bigg \}
\end{align}
is minimized with respect to the variational parameters. The time-dependent Lagrange multipliers are introduced to ensure the orbitals to remain orthonormal during the temporal evolution.

We first derive the equations of motion for the orbitals by requiring the variation to be stationary,
\begin{align}
\frac{\delta}{\delta \smallbra{\phi_n}} \; S\Bigl[\bigl \{C_{\mathbf n}(t) \bigr \}, \bigl \{\smallket{\phi_k}(t)\bigr\} \Bigr] \ \stackrel{!}{=} \ 0 \,.
\end{align}
After a few steps we arrive at the following nonlinear equation:
\begin{align}
\label{MCTDH_orbital_1} \hat{\mathbf  P} \; i \tderiv \; \ket{ \phi_n} \ = \ & \hat{\mathbf  P} \; \Biggl\{ \, \hat h(t) \; \ket{ \phi_n} \; + \; \sum_{pqrs} \; \left( {\mathbf D}^{-1}\right)_{np} \; d_{pqrs} \; \hat g_{rs} \; \ket{ \phi_q} \; \Biggr\}\,,
\end{align}
where we introduced the (spin-restricted) one- and two-particle density matrices
\begin{align}
D_{pq} \ &= \  \braket{\Psi}{\hat E_{pq}}{\Psi}\,, \\
d_{pqrs} \ &= \ \braket{\Psi}{\hat e_{pqrs}}{\Psi}\,,
\end{align}
as well as the mean-field operator $\hat g_{rs}$, which in coordinate representation reads
\begin{align}
g_{rs}(\mathbf r) \ = \ \int d\mathbf r^\prime \; \phi^\ast_r(\mathbf r^\prime) \, \frac{1}{|\mathbf r -\mathbf r^\prime|} \, \phi_s(\mathbf r^\prime)\,.
\end{align}
Further, the elimination of the Lagrange multipliers led to a projection operator
\begin{align}
\hat{\mathbf P} \ = \ 1 \, - \, \sum_m \, \ket{\phi_m}\bra{\phi_m}\,,
\end{align}
which projects on the orthogonal complement of the span of the orbitals.
In order to remove the projection operator on the lhs.~of Eq.~(\ref{MCTDH_orbital_1}) and obtain explicit equations, a unitary transformation among the orbitals is applied, which ensures
\begin{align}
\label{unitary_trans} \bigbraket{\phi_k}{\tderiv}{\phi_l} \ = \ 0\,,
\end{align}
i.e. the change of an orbital is orthogonal to the subspace spanned by all orbitals. After insertion into Eq.~(\ref{MCTDH_orbital_1}), we obtain the MCTDH orbital equations
\begin{align}
\label{MCTDH_orbital} i \tderiv \; \ket{ \phi_n} \ = \ & \hat{\mathbf  P} \; \Biggl\{ \, \hat h(t) \; \ket{ \phi_n} \; + \; \sum_{pqrs} \; \left( {\mathbf D}^{-1}\right)_{np} \; d_{pqrs} \; \hat g_{rs} \; \ket{ \phi_q} \; \Biggr\}\,.
\end{align}

The minimization with respect to the coefficients, i.e.
\begin{align}
\frac{\delta}{\delta C^\ast_{\mathbf n}}\; S\Bigl[\bigl \{C_{\mathbf n}(t) \bigr \}, \bigl \{\smallket{\phi_k}(t)\bigr\} \Bigr] \ \stackrel{!}{=} \ 0\,,
\end{align}
then straightforwardly leads to a Schrödinger equation in matrix representation in the SD-basis:
\begin{align}
\label{MCTDH_coefficient} i \tderiv \, C_{\mathbf n}(t) \ = \ \sum_{\mathbf m} \braket{\mathbf n}{ \hat H(t) }{\mathbf m} \; C_{\mathbf m}(t)\,.
\end{align}
Note, that again Eq.~(\ref{unitary_trans}) has been used, which also causes the matrix element of the time-derivative operator between Slater determinants to vanish,
\begin{align}
\bigbraket{\mathbf n}{ \tderiv }{\mathbf m} \ = \ 0\,.
\end{align}

\begin{figure}[t]
  \begin{center}
    \includegraphics[width=0.65\textwidth]{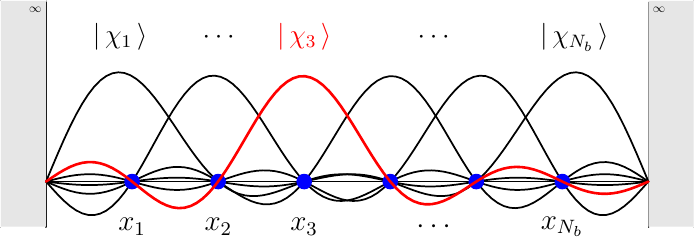}
  \end{center}
  \caption{\label{fig:sine_DVR}Schematic view of the sine DVR functions, for a basis size of \mbox{$N_b=6$}. The basisfunctions are constructed over an equidistant grid $x_k$ in such a way, that \mbox{$\chi_i(x_k)\,=\, \delta_{ik} / \sqrt{w_k}$} holds, with a set of integration weights $w_k$. Matrix elements may then be evaluated approximately by a summation, i.e. $\int dx  \, \chi^\ast_i(x) f(x) \chi_j(x) \ \longrightarrow \ \sum_k w_k \, \chi^\ast_i(x_k) f(x_k) \chi_j(x_k)$. This leads directly to diagonal spatial matrix elements, $\braket{\chi_i}{ f(\hat x)}{\chi_j} = f(x_i) \, \delta_{ij}$.}
\end{figure}

\subsection{Numerical implementation}\label{sec:numerical}
We give some notes on our numerical solution of the coupled set of MCTDHF equations, Eqs.~(\ref{MCTDH_coefficient}) and (\ref{MCTDH_orbital}).
\\
\emph{Single-particle basis.} First of all, like in Eq.~(\ref{basis_ex}), an appropriate time-independent single particle basis $\{ \smallket{\chi_l} \}$ is chosen, which inserted into the orbital equation (\ref{MCTDH_orbital}) yields an equation for the time-dependent expansion coefficients $b_{kl}(t)$ \cite{Hochstuhl_2010}. In this work we use a sine discrete variable representation (DVR) basis \cite{Colbert_1992, Beck_1997}, see Fig.~\ref{fig:sine_DVR}. As is common to all DVR bases, matrix elements of spatial operators are diagonal, and they are simply given by 
the function values on a grid $\mathbf r_i$ of Gaussian integration points,
\begin{align}
\braket{\chi_p}{f(\mathbf r)}{\chi_q} \ &= \ \delta_{pq} \; f(\mathbf r_q) \,,\\
\braket{\chi_p \, \chi_r }{g(\mathbf r,{\mathbf r}^\prime)}{\chi_s \, \chi_q} \ &= \ \delta_{pq} \; \delta_{rs} \; g(\mathbf r_i, \mathbf r_j)\,.
\end{align}
For the sine DVR, the grid consists of equally spaced nodes and the non-diagonal kinetic energy matrix can be evaluated analytically.\footnote{Instead of a normal DVR, one could also employ a finite-element DVR, see e.g. \cite{Balzer_2009,Balzer_Glasgow_2009}.}
\\
\emph{Density matrices and Hamiltonian.} For the evaluation of the density matrices and the Hamiltonian, the matrix elements of one-and two-particle excitation operators in the basis of Slater determinants, $\braket{\mathbf n}{\hat E_{pq}}{\mathbf m}$ and $\braket{\mathbf n}{\hat e_{pqrs}}{\mathbf m}$, have to be evaluated, both of which may attain either zero, plus or minus one. Since these quantities are needed very frequently, all non-zero contributions are determined and their sign is stored in memory before the actual time propagation. The Hamiltonian can then be easily calculated using the Slater-Condon rules \cite{Helgaker}.
\\
\emph{Time evolution.} After writing the wavefunction coefficients $\mathbf C$ and the single-particle basis expansion coefficients $\mathbf b$ in a single vector, the MCTDHF equations may be casted to the general form
\begin{align}
\label{MCTDH_vector} i\, \tderiv \; \boldsymbol {\mathcal V} \ = \ \boldsymbol {\mathcal F}(\boldsymbol {\mathcal V}) \ , \qquad \boldsymbol {\mathcal V} \ = \ \begin{pmatrix} \, {\mathbf b} \, \\ \, {\mathbf C} \, \end{pmatrix}\,.
\end{align}
We solve this coupled set of equations by means of general purpose integrators, such as Runge-Kutta or Burlisch-Stoer methods \cite{numerical_recipes}. Other possible techniques particularly well suited for the MCTDHF scheme are given in Ref. \cite{Beck_1997}.
\\
\emph{Solution effort.} As mentioned above, the basis of Slater-determinants grows with $\binom {2M}N$, where $M$ is the number of time-dependent spatial orbitals and $N$ the number of particles, leading to the typical exponential problem of Configuration Interaction. However, for the two-particle model we consider here, the SD-basis only grows with $\mathcal O(M^2)$ and thus poses no difficulties. Here, the effort is rather determined by the number $N_b$ of orbitals, which may become large in order to adequately describe the continuum. For each evaluation of the rhs. $\boldsymbol {\mathcal F}(\boldsymbol {\mathcal V})$ of Eq.~(\ref{MCTDH_vector}), the electron integrals in the time-dependent basis are needed, which are formed through a transformation from the time-independent basis using the orbital coefficients. The time-consuming part is the transformation of the two-particle interaction matrix. Assuming $N_b \gg M$, the leading term for a DVR basis is given by $\mathcal O(M N_b^{\,2})$. To further diminish the effort and obtain an almost linear scaling with the size of the underlying time-independent basis, low-rank approximations of the interaction potential can be used \cite{Caillat_2005,Beck_1997}. We also note, that for a fixed basis size $N_b$ the effort grows as $\mathcal{O}(M^4 N_b)$.

\begin{table}[b]
 \begin {center}
  \begin{tabular}{| c | c | c | c | c | c | c |}
  \hline
  \parbox[0pt][1.6em][c]{0cm}{} $\phantom{xxx} M \phantom{xxx}$ &  $\phantom{xxx} 1 \phantom{xxx}$ & $\phantom{xxx} 2 \phantom{xxx}$ & $\phantom{xxx} 3 \phantom{xxx}$ & $\phantom{xxx} 4 \phantom{xxx}$ & $\phantom{xxx} 5 \phantom{xxx}$& $\phantom{x} \text{exact} \phantom{x}$\\
  \hline
  \parbox[0pt][1.6em][c]{0cm}{} Energy [Hartree]& $-2.2242$ & $-2.2365$ & $-2.2381$ & $-2.2382$ & $-2.23825$ & $-2.23826$\\
  \parbox[0pt][1.6em][c]{0cm}{} \% of corr. energy & $-$ & $87 \%$ & $98.9 \%$ & $99.6 \%$ & $99.9 \%$ & $100 \%$\\
  \hline
  \end{tabular}
 \end{center}
 \caption{\label{tab:energies}Total energies of the groundstate for different numbers $M$ of time-dependent orbitals, plus the fraction of the included correlation energy (the entire correlation energy is defined as the difference between the exact and the Hartree-Fock result). With increasing $M$, the energy rapidly approaches the exact energy as obtained by a direct solution of the Schrödinger equation. For $M>5$, the results equal the exact result for the given number of digits.}
\end{table}

\begin{figure}[t]
  \begin{center}
    \includegraphics[width=0.9\textwidth]{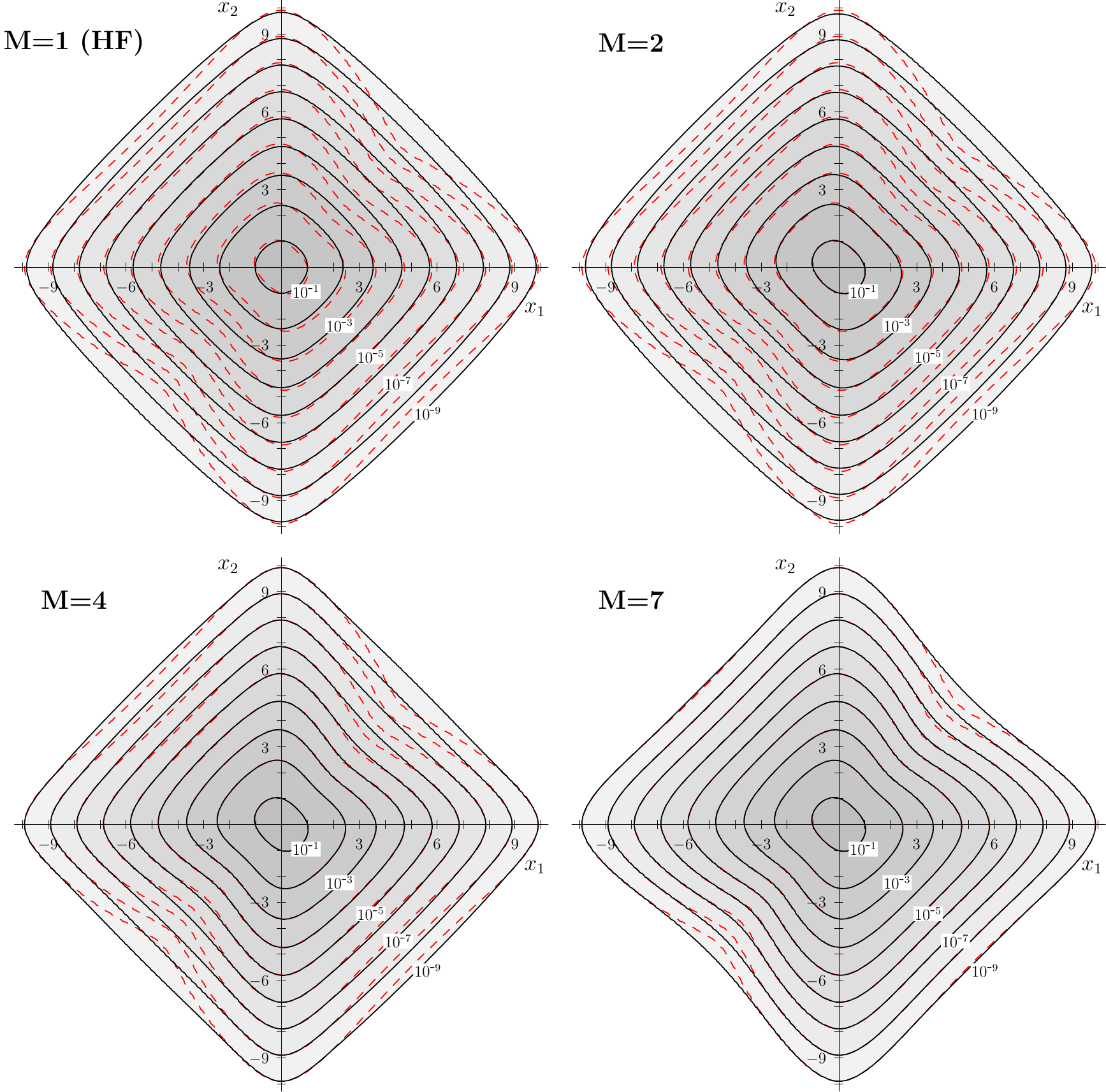}
  \end{center}
  \caption{\label{fig:densities} Logarithmic contour plot of the groundstate two-particle densities for different numbers $M$ of MCTDHF orbitals. $x_1$ and $x_2$ are the coordinates of the two electrons. In each plot, the red dashed curves show the result from a direct solution of the Schrödinger equation.}
\end{figure}

\section{Numerical results}
\subsection{One-dimensional Helium model}\label{sec:helium_model}
The here considered one-dimensional model of Helium is given by a potential
\begin{align}
V(x) \ = \ -\frac{2}{\sqrt{x^2+1}}\,,
\end{align}
and has been well tested for roughly 30 years. Compared to the real three-dimensional Helium atom, in this model the electron movement is restricted only to the laser polarization axis.
Its usefulness derives on the one hand from the fact, that electronic correlation effects of Helium and also larger atoms can be qualitatively well explained, most prominently the nonsequential double ionization \cite{Liu_1999,Dahlen_2001}. On the other hand, it is exactly solvable by viewing the one-dimensional two-particle problem as a Schrödinger equation of one-particle moving in the two-dimensional external potential
\begin{align}
V(x,y) \ = \ -\frac{2}{\sqrt{x^2+1}} -\frac{2}{\sqrt{y^2+1}} +\frac{1}{\sqrt{(x-y)^2+1}}\,.
\end{align}
In this paper, we use it as a benchmark for the approximate MCTDHF simulations. For the details of the employed solution method of the two-dimensional TDSE, see \cite{Bauch_2008}.

\subsection{Groundstate results}
In the following we present the results for the groundstate of the Helium model obtained with MCTDHF. A similar investigation has already been given in \cite{Zanghellini_2004}, however with a slightly different atomic potential. The groundstate is obtained by propagation of the MCTDHF equations in imaginary time (ITP), starting from the orbitals of the noninteracting system. The system is evolved until the difference of the total energy between two steps falls below a certain limit (here $10^{-10} \, \text{a.u.}$). The results are compared to direct solutions of the Schrödinger equation, which were obtained by ITP as well. In Tab.~\ref{tab:energies} we display the groundstate energies for different MCTDHF approximations, and the corresponding fraction of the correlation energy as defined by the difference between the exact and the Hartree-Fock ($M=1$) energy. We observe a rapid convergence with the number of orbitals, already the first correction to Hartree-Fock, $M=2$, is able to account for $87 \%$ of the entire correlation energy. Note that $M=2$ only needs roughly an eight-fold effort as compared to Hartree-Fock, which is due to the CPU time growth with $\mathcal{O}(M^4N_b )$ (see section \ref{sec:numerical}).\\
In Fig.~\ref{fig:densities} we plot the two-particle (tp) densities, that for a two-electron system equal the absolute square of the wavefunction, again for different approximations. In each plot, the (red) dashed curve depicts the tp-density from the exact solution. From the figure it is obvious, that the Hartree-Fock approximation is not able to reproduce the correct butterfly shape caused by the Coulomb repulsion (which is largest for $x_1=x_2$). With increasing number of orbitals, the tp-density approaches the exact result, until it attains an accuracy of seven digits for $M=7$.

\begin{figure}[t]
  \begin{center}
    \includegraphics[width=1.0\textwidth]{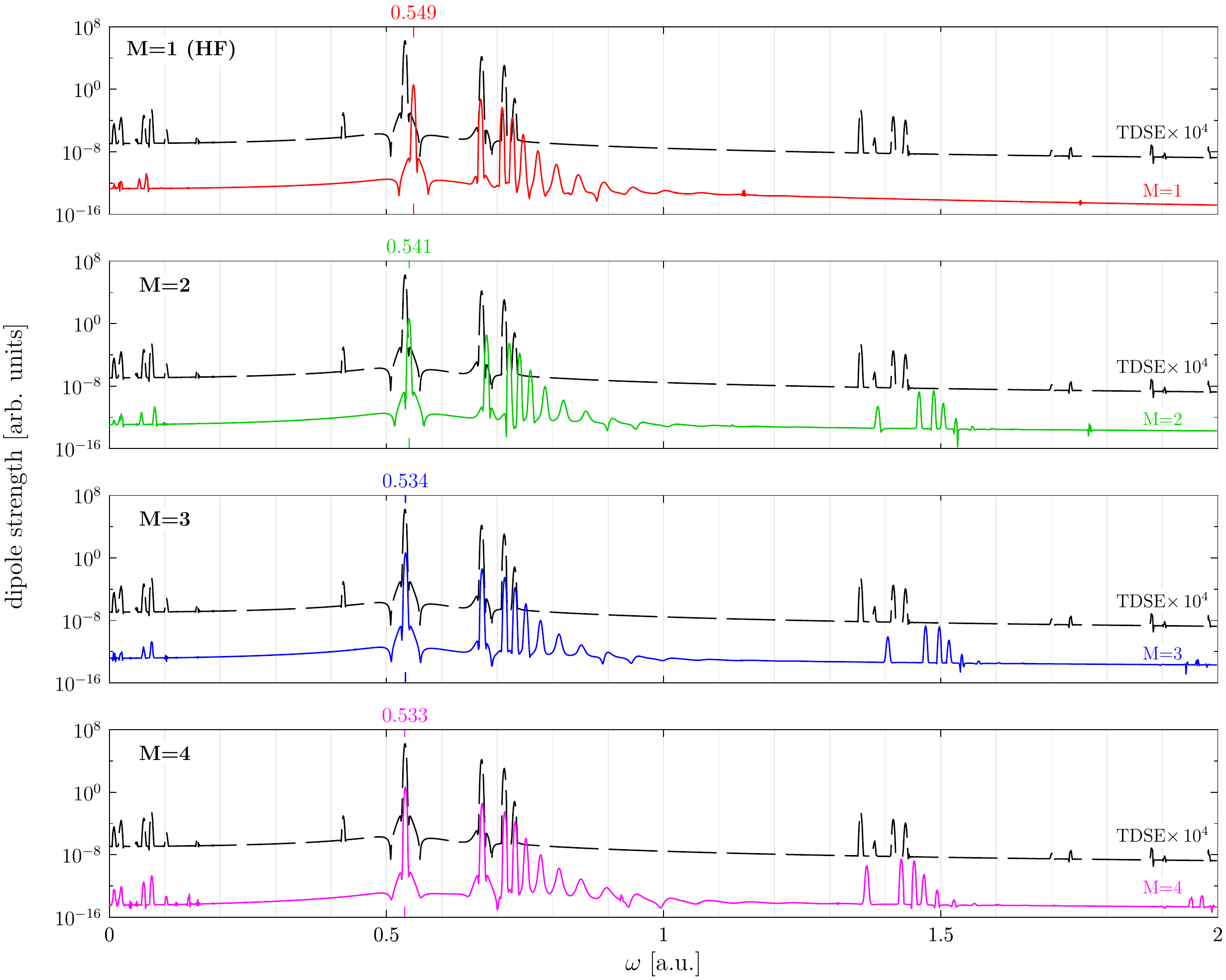}
  \end{center}
  \caption{\label{fig:dipole}Linear response of the one-dimensional Helium model atom as obtained from MCTDHF and the TDSE (shifted by $10^4$). The pictures show results for different numbers $M$ of time-dependent orbitals. The peaks in the tail of the spectrum are caused by correlations. The numbers on the upper $x$-axes denote the positions of the first peak, which with increasing $M$ converge towards the exact position $\omega = 0.533\, \text{a.u.}$}
\end{figure}

\subsection{Time-dependent results}
As an example for time-dependent calculations, we investigate the reaction of the model atom to an external perturbation in linear response\footnote{The linear response can also be obtained from time-independent calculations by diagonalization of the Hamiltonian. In fact, the following results present an alternative for the diagonalization called ``filter-diagonalization'', see e.g. \cite{Mandelshtam_1997}.}. To this end, we disturb the groundstate by a dipole kick, which is sufficiently short in order to provide a homogeneous spectral density and sufficiently weak to avoid non-linear effects. In this work, we chose a kick with duration $0.01 \, \text{a.u.}$ and an amplitude of $\mathcal{E}_0=0.01 \, \text{a.u.}$ The disturbed system was propagated for $2000 \,\text{a.u.}$ ($\sim 50 \, \text{fs}$) and the expectation value of position has subsequently been Fourier-transformed (using a Blackman-Harris window). This yields the dipole spectrum, from which the excitation frequencies can be observed. In Fig.~\ref{fig:dipole} we plot the obtained spectra for different approximations and compare them with the result from the TDSE. First let us consider the energy region below the first ionization threshold $\omega\leq 0.75 \, \text{a.u.}$ \cite{Haan_1994}, which corresponds to the one-electron excitations. These excitations are rather well reproduced already within Hartree-Fock, only the position of the peaks and thus the excitation energies deviate slightly. However, for Hartree-Fock ($M=1$), there are no peaks in the tail of the curve, which consequently are caused by correlations. Already in the first correlation correction, $M=2$, the main peaks exist, but their positions deviate from the TDSE result. For $M=4$, the exact spectrum is well reproduced. The values on the upper $x$-axes indicate the position of the first peak, which determine the difference between the groundstate energy and the energy of the first excited singlet state. We observe a similar trend as before, namely that $M=4$ is able to reproduce the exact result of $\omega=0.533 \, \text{a.u.}$

\section{Summary}
We have given an introduction to the time-dependent Multiconfiguration Hartree-Fock formalism and discussed the main ideas of our numerical implementation. In order to investigate the characteristics and capabilities of the method, it has been applied to a well-known one-dimensional model of the Helium model. We calculated the groundstate energies and two-particle densities, as well as the linear response of the system, each one for different MCTDHF approximations, i.e. for different numbers $M$ of time-dependent spatial orbitals. All results were compared to direct solutions of the time-dependent Schrödinger equation. Our investigation showed, that MCTDHF is well suited for the time-resolved study of this two-electron system. We are very confident, that a comparable level of accuracy can also be obtained for larger system (up to, say, $N=10$).\\
Our future work therefore will concentrate on time-dependent electronic correlations in larger model atoms, which are not accessible by direct solutions of the Schrödinger equation.

\section*{Acknowledgements}
This work is supported by the Deutsche Forschungsgemeinschaft via SFB-TR 24 and the U.S. Department of Energy award DE-FG02-07ER54946.

\bibliography{Proc_Glasgow}

\end{document}